\documentstyle[aps,prl,psfig]{revtex}

\def\be{\begin{equation}}           
\def\ee{\end{equation}}
\def\lsim{\lower0.5ex\hbox{$\; \buildrel < \over \sim \;$}}
\def\ergs{\mbox{erg\,s$^{-1}$}}
\def\gcm{\mbox{G\,cm$^3$}}
\def\ergcms{\mbox{erg\,cm$^{-2}$\,s$^{-1}$}}
\def\dm{\mbox{$\dot{M}$}}
\def\msun{\mbox{$M_{\odot}$}}
\def\ra{\mbox{$R_{\rm A}$}}
\def\rco{\mbox{$R_{\rm c}$}}
\def\ro{\mbox{$R_0$}}
\def\dmmax{\mbox{$\dm_{\rm max}$}}
\def\dmmin{\mbox{$\dm_{\rm min}$}}
\def\fmax{\mbox{$F_{\rm max}$}}
\def\fmin{\mbox{$F_{\rm min}$}}

\def\apj{Astrophys. J.\ }
\def\aap{Astron. Astrophys.\ }
\def\nat{\mbox{Nature\ }}
\def\sci{\mbox{Science\ }}
\def\sax{\mbox{SAX J1808.4-3658}}
\def\etal{\mbox{\it et al.}}

\begin{document}

\draft

\title{Is SAX J1808.4-3658 A Strange Star?}

\author{X.-D. Li}
\address{
Department of Astronomy, Nanjing University, Nanjing 210093, China}
\author{I. Bombaci}
\address{Dipartimento di Fisica, Universit\'{a} di Pisa
and I.N.F.N. Sezione di Pisa, via Buonarroti 2, I-56100 Pisa, Italy}
\author{M. Dey}
\address{Department of Physics, Presidency College, Calcutta 700 073, India,
and an associate member of the International Centre for Theoretical Physics,
Trieste, Italy}
\author{J. Dey}
\address{IFT-UNESP, 145 Rua Pamplona,  Sao Paulo 01405-900, Brasil and Azad
Physics Centre, Maulana Azad College, Calcutta 700 013, India}
\author{E. P. J. van den Heuvel}
\address{Astronomical Institute, University of Amsterdam, Kruislaan 403, 
1098 SJ Amsterdam, The Netherlands}

\date{\today}
\maketitle

\begin{abstract}

One of the most important questions in the study of compact objects is the
nature of pulsars, including whether they are composed of $\beta$-stable
nuclear matter or strange quark matter.  Observations of the newly
discovered millisecond X-ray pulsar \sax\ with the Rossi X-Ray Timing
Explorer place firm constraint on the radius of the compact star.
Comparing the mass - radius relation of \sax\ with the theoretical
mass - radius relation for neutron stars and for strange stars, we find
that a strange star model is more consistent with \sax, and suggest that
it is a likely strange star candidate.

\end{abstract}
\pacs{PACS numbers: 98.70.Rz,12.38.Mh,26.60.+c,97.60.Jd  \hfill}

The transient X-ray burst source \sax\ was discovered in September 1996 with
the Wide Field Camera (WFC) on board BeppoSAX \cite{in98}. Two bright type I 
X-ray bursts were detected, each lasting less than 30 seconds. Such bursts
are generally accepted to be due to thermonuclear flashes on the surface of
a neutron star \cite{lpt95}, suggesting that this source is a member of 
low-mass X-ray binaries (LMXBs), consisting of a low ($\lsim 10^{10}$ G) 
magnetic field neutron star accreting from a companion star of less than one 
solar mass \cite{bh91}.
Analysis of the bursts in \sax\ indicates that it is 4 kpc distant and has
a peak X-ray luminosity of $6\times 10^{36}\,\ergs$ in its bright state,
and $<10^{35}\,\ergs$ in quiescence \cite{in98}.
 
Recently a transient X-ray source designated XTE J1808-369 was detected with
the Proportional Counter Array (PCA) on board the Rossi X-ray Timing
Explorer (RXTE) \cite{m98}. The source is positionally coincident within a 
few arcminutes with \sax, implying that both sources are the same object.
Coherent pulsations at a period of 2.49 milliseconds were discovered 
\cite{wk98}.  The star's surface dipolar magnetic moment was derived to be 
$\lsim 10^{26}\,\gcm$ from detection of X-ray pulsations at a luminosity of 
$10^{36}\,\ergs$ \cite{wk98}, consistent with the weak fields expected for 
type I X-ray bursters \cite{lpt95} and millisecond radio pulsars (MS PSRs) 
\cite{bh91}. The binary nature of \sax\ was firmly established with the 
detection of a 2 hour orbital period \cite{cm98}, as well as with the optical 
identification of the companion star \cite{r98}. \sax\ is the first pulsar to 
show both coherent pulsations in its persistent emission and thermonuclear 
bursts, and by far the fastest-rotating, lowest-field accretion-driven pulsar 
known.  It presents direct evidence for the evolutionary link between LMXBs 
and MS PSRs \cite{bh91}.
 
The discovery of \sax\ also allows a direct test of the compactness of 
pulsars. Detection of X-ray pulsations requires that the inner radius $\ro$ 
of the accretion flow (generally in the form of a Keplerian accretion disk 
in LMXBs) should be larger than the stellar radius $R$ (viz.\,the stellar 
magnetic field must be strong enough to disrupt the disk flow above the 
stellar surface), and less than the so-called corotation radius 
$\rco=[GM/(4\pi^2)P^2]^{1/3}$ (viz.\,the stellar magnetic field must be weak 
enough that accretion is not centrifugally inhibited) \cite{bk98,pc98}. Here 
$G$ is the gravitation constant, 
$M$ is the mass of the star, and $P$ is the pulse period. The inner disk 
radius $\ro$ is generally evaluated in terms of the Alfv\'en radius $\ra$, 
at which the magnetic and material stresses balance \cite{bh91},
$\ro=\xi\ra=\xi[B^2R^6/\dm(2GM)^{1/2}]^{2/7}$, where $B$ and $\dm$ are
respectively the surface magnetic field and the mass accretion 
rate of the pulsar, and $\xi$ is a parameter of order of unity almost 
independent of $\dm$ \cite{bk98,l97}.
Since X-ray pulsations in \sax\ were detected over a wide range of mass 
accretion rate, say, from $\dmmin$ to $\dmmax$, a firm upper limit of the 
stellar radius can then be obtained from the condition 
$R<\ro(\dmmax)<\ro(\dmmin)<\rco$, i.e.,
\be
R<27.6 (\frac{\fmax}{\fmin})^{-2/7}(\frac{P}{2.49\,{\rm ms}})^{2/3}
   (\frac{M}{\msun})^{1/3}\,{\rm km},
\ee
where $\fmax$ and $\fmin$ denote the X-ray fluxes measured during X-ray
high- and low-state, respectly, $\msun$ is the solar mass. Here we have 
assumed that the mass accretion rate $\dm$ is proportional to the X-ray flux 
observed with RXTE. This is guaranteed by the fact that the X-ray spectrum 
of \sax\ was remarkably stable \cite{g98} and there was only slightly increase 
in the pulse amplitude \cite{cmt98} when the X-ray luminosity varied by
a factor of $\sim 100$ during the 1998 April/May outburst.

\begin{figure}
\centerline{\psfig{figure=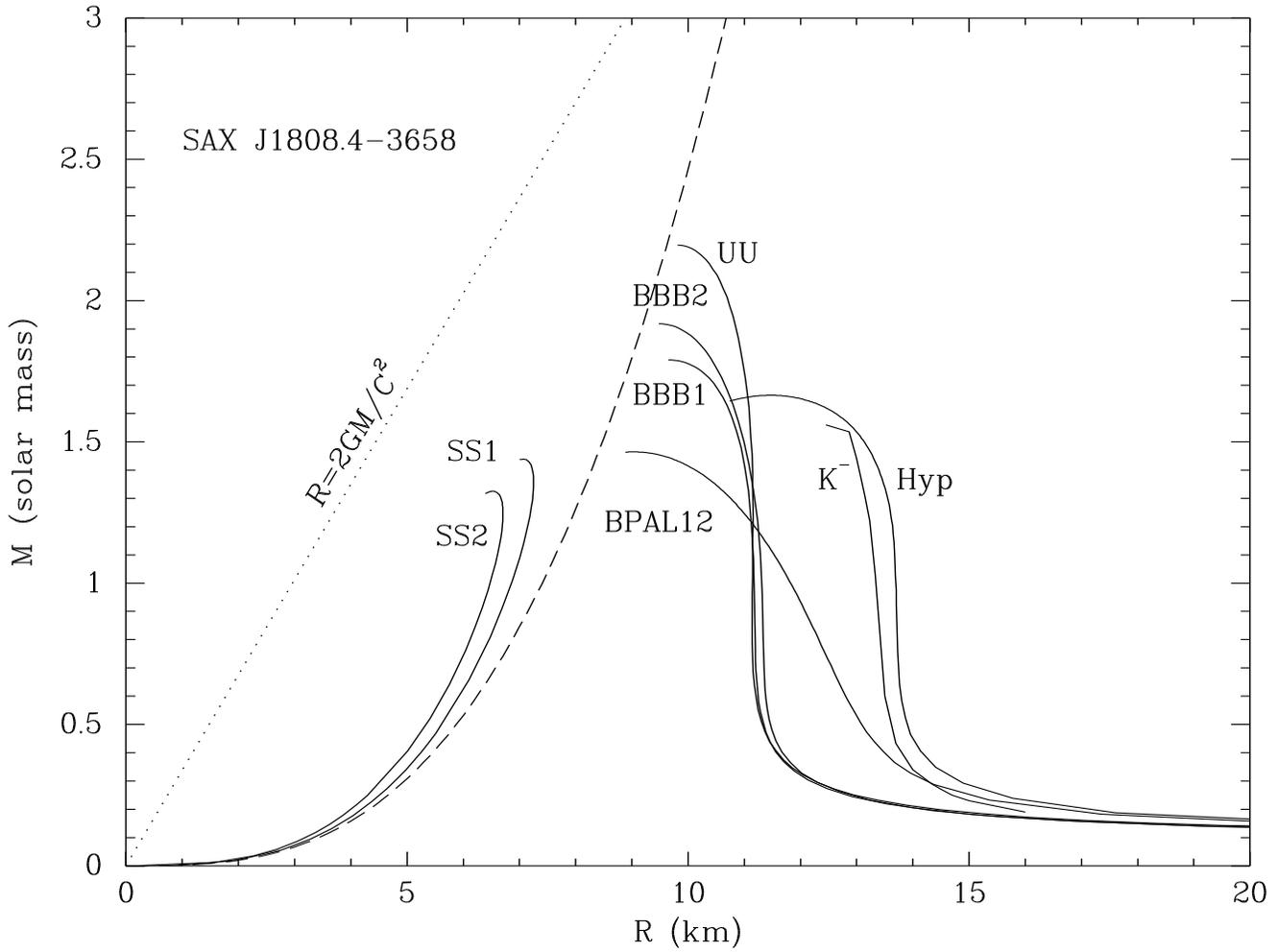,angle=-90}}
\caption{
Comparison of the $M-R$ relation of \sax\ determined from RXTE observations 
with theoretical models of neutron stars and of strange stars.  The range of
mass and radius of \sax\ is allowed in the region outlined by the dashed and 
dotted curves. The solid curves labeled UU, BBB1, BBB2, BPAL12, hyp, and 
$K^-$ represent various $M-R$ relations for {\it realistic} EOSs of nonrotating 
neutron stars; the solid curves labeled ss1 and ss2 are for strange 
stars (see text for details and references to the EOS models).  
It is obvious that strange stars are more consistent with the \sax\ 
properties compared to neutron stars.
}
\label{fig}
\end{figure}

Given the range of X-ray flux at which coherent pulsations were detected,
inequality (1) defines a limiting curve in the mass - radius ($M-R$) parameter
space for \sax, as plotted in the dashed curve in Fig.\ \ref{fig}. 
During the 1998 April/May outburst, the maximum observed $2-30$ keV flux of 
\sax\ at the peak of the outburst was $\fmax\simeq 3\times 10^{-9}\,\ergcms$, 
while the pulse signal became barely detectable when the flux dropped below 
$\fmin\simeq 2\times 10^{-11}\,\ergcms$ \cite{cmt98}. Here we adopt 
$\fmax/\fmin\simeq 100$. The dotted curve represents the Schwarzschild 
radius $R=2GM/c^2$ (where $c$ is the speed of light) - the lower limit of 
the stellar radius to prevent the star collapsing to be a black hole 
\cite{st83}. Thus the allowed range of the mass and radius of \sax\ is the 
region confined by the dashed and dotted curves in Fig.\ \ref{fig}.
 
Figure\ \ref{fig} compares the theoretical $M-R$ relations (solid curves) 
for nonrotating neutron stars given by six recent realistic models for the 
equation of state (EOS) of dense matter. In models UU \cite{wff88}, BBB1 and 
BBB2 \cite{bbb97} the neutron star core is assumed to be composed by an 
uncharged mixture of neutrons, protons, electrons and muons in equilibrium 
with respect to the weak interaction ($\beta$-stable nuclear matter). 
Equations of state UU, BBB1, BBB2 are based on microscopic calculations of 
asymmetric nuclear matter by use of realistic nuclear forces which fit 
experimental nucleon-nucleon scattering data, and deuteron properties.   
In model Hyp \cite{p97}, hyperons are considered in addition to nucleons as 
hadronic constituents of the neutron star core. Next, we consider, as a 
limiting case, a very {\it soft} EOS for $\beta$-stable nuclear matter, 
namely the BPAL12 model \cite{p97}, which is still able to sustain the 
measured mass 1.442 $M_\odot$ of the pulsar PSR 1916+13. In general, a 
{\it soft} EOS is expected to give a lower limiting mass and a smaller radius 
with respect to a {\it stiff} EOS \cite{st83}.  
Finally, we consider the possibility that neutron stars may possess 
a core with a Bose--Einstein condensate of negative kaons~\cite{KN,kyo1,tpl}. 
The main physical effect of the onset of $K^-$ condensation is a softening 
of the EOS with a consequent lowering of the neutron star maximum mass   
and possibly of the radius.  
Actually, neutron star with $R \sim 7-9~km$ were obtained~\cite{kyo1,tpl},  
for some EOS with $K^-$ condensation.  
However, in those models~\cite{kyo1,tpl} the kaon condensation phase 
transition was implemented using the Maxwell construction, which is inadequate 
in stellar matter, where one has two conserved charges:  
baryon number and electric charge~\cite{glend}.         
When the kaon condensation phase transition is implemented 
properly~\cite{glend}, one obtains neutron stars with ``large'' radii, 
as shown by the curve labeled $K^-$ in Fig.\ \ref{fig}. 
Moreover, kaon-nucleon and nucleon-nucleon correlations rise the threshold 
density for the onset of kaon condensation, possibly to densities higher than 
those found in the centre of stable neutron stars~\cite{ppt}.  
It is clearly seen in Fig.\ \ref{fig} that none of the neutron star $M-R$ 
curves is consistent with \sax\ (Including rotational effects will shift  
the $M-R$ curves to up-right in Fig.\ \ref{fig} \cite{dtb98}, and does not 
help improve the consistency 
between the theoretical neutron star models and observations of \sax).  
Moreover, it is unlikely that the actual mass and radius of \sax\ lie very
close to the dashed curve, since the minimum flux $\fmin$ at which X-ray
pulsations were detected by RXTE was determined by the instrumental
sensitivity, and the actual value could be even lower; while the presence
of the slight X-ray dips observed in \sax \cite{cm98} suggests that the 
companion mass is most likely to be less than $0.1\msun$, and the pulsar 
mass op more than $1\msun$. Therefore it seems that \sax\ is not well 
described by a neutron star model. As shown below, a strange star model 
seems to be more compatible with \sax.

Note that in writing inequality (1) we have implicitly assumed that 
the pulsar magnetic field is basically dipolar, even when the accretion 
disk is close to the stellar surface. 
This is partly supported by the agreement between the dipolar
spin-up line and the location of MS PSRs in the spin period - spin period
derivative diagram, which implies that the multipole moments in LMXBs are
no more than $\sim 40\%$ of the dipole moments if the quadrupole component
is comparable to or larger than higher order anomalies \cite{a93}.
However, the $\ro(\dm)$ relation will be changed if the star's field has
more complicated structure. For example, there may exist regions on the
surface of the star where the magnetic field strength is much greater than
that from a central dipole, to affect the channelling of the accretion 
flow and the pulsed emission; the induced current flow in the boundary 
layer of the disk could increase the field strength when $\ro$ reaches $R$.
If \sax\ possesses such anomalous field, there should be
two kinds of observational effects: the pulse profile shows a
dependence on energy (because in strongly magnetized plasma, photons with
different energies have different scattering cross-sections), and the
X-ray spectrum changes with the mass accretion rate (due to a change in
the configuration of accretion pattern and in the X-ray emitting region).
These are in contrast with the observations of \sax, which shows a single
sine pulse profile with little energy dependence \cite{cmt98}, and stable 
X-ray spectra when the X-ray luminosity varied by a factor of $\sim 100$ 
\cite{g98}. Note also that in the $\ro(\dm)$ formula the parameter 
$\xi\sim \ro/h$, where $h$ is the scale height of the disk \cite{wy}. 
The effect of the increase in $B_r$ with $\dm$, due to the induced current,
may be largely counteracted by the decrease of $\xi$ with $\dm$, and a 
steeper $\dot{M}$-dependence of $\ro$ when the effect of general relativity 
is included \cite{l98}. So we conclude that the accretion flow around \sax\ 
may still be dominated by a central dipole field even when the
disk reaches the star. 

Strange stars are astrophysical compact objects which are entirely made of 
deconfined {\it u,d,s} quark matter ({\it strange matter}).  The possible 
existence of strange stars is a direct consequence of the conjecture 
\cite{bw} that strange matter may be the absolute ground state of strongly 
interacting matter.  Detailed studies have shown that the existence of
strange matter is allowable within uncertainties inherent in a strong  
interaction calculation \cite{fj84}; thus strange stars may exist in the 
universe.  Apart from the fact that strange stars may be relics from the
cosmic separation of phases as suggested by Witten \cite{bw}, a seed
of strange matter may convert a neutron star to a strange one \cite{o87}.
Conversion from protoneutron stars during the collapse of supernova cores is 
also possible \cite{g93}.  
Recent studies have shown that the compact objects associated with the X-ray 
pulsar Her X-1 \cite{ldw95,d98}, and with the X-ray burster 
4U 1820-30 \cite{b97}, are good strange star candidates.
 
Most of the previous calculations~\cite{afo86} of strange star properties 
used an EOS for strange matter based on the phenomenological nucleonic bag 
model, in which the basic features of quantum chromodynamics, such as quark 
confinement and asymptotic freedom are postulated from the beginning. 
The deconfinement of quarks at high density is, however, not obvious in the 
bag model. To find a star of small mass and radius, one has to postulate a 
large bag constant, whereas one would imagine in a high density system the 
bag constant should be lower.

Recently, Dey \etal \cite{d98} derived an EOS for strange matter, which has 
asymptotic freedom built in, shows confinement at zero baryon density,  
deconfinement at high density, and gives a stable configuration for 
chargeless, $\beta$-stable strange matter.
In this model the quark interaction is described by  an interquark vector 
potential originating from gluon exchange, and by a density dependent scalar 
potential which restores the chiral symmetry at high density. This EOS was 
then used \cite{d98} to calculate the structure of strange stars.  
Using the same model (but different values of the parameters with respect to 
those employed in ref.~\cite{d98}) we calculated the $M-R$ relations, 
which are also shown in solid curves labeled ss1 
and  ss2  in Fig.\ \ref{fig}, corresponding to strange stars with 
maximum masses of $1.44 \,\msun$ and $1.32\,\msun$ \cite{Mmax} 
and radii of 7.07 km and  6.53 km, respectively.  
It is seen that the region confined by the dashed and dotted curves 
in Fig.\ \ref{fig} is in remarkable accord with the strange 
star models. Figure \ref{fig} clearly demonstrates that a strange star model 
is more compatible with \sax\ than a neutron star one.  

If \sax\ is a strange star, then the thermonuclear flash model \cite{lpt95} 
can not be invoked to explain the observed X-ray bursts.   
However, a different mechanism has been recently proposed \cite{c98}, in which 
the X-ray burst is powered basically by the conversion of the accreted normal 
matter to strange quark matter on the surface of a strange star. 

As both the spin rate and the magnetic moment of \sax\ resemble those 
inferred for other, non-pulsing LMXBs, an interesting and important question 
is: why is \sax\ the only known LMXB with an MS PSR? The most straightforward 
explanation seems to be that the magnetic field of \sax\ is considerably
stronger than that of other systems of similar X-ray luminosity \cite{wk98}.
We point out that a strange star is more liable to radiate pulsed emission 
than a neutron star because of its compactness. As seen in Fig.\ \ref{fig}, 
the radius of a $\sim 1\,\msun$ strange star is generally $1.5-2$ times 
smaller than that of a neutron star of similar mass, implying that, with the 
same magnetic moment (the observable quantity), the surface field strength of 
the strange star is $3-8$ times higher than that of the neutron star,  
and that the size of the polar caps in the strange star for field-aligned 
flow, $4\pi R^2(1-(1-R/\ro)^{1/2})$, is up to 10 times smaller than in the 
neutron star.  The more efficient magnetic 
channelling of the accreting matter close to the strange star surface could 
then lead to higher pulsation amplitudes, making it easier to detect.
A strange star model for \sax\ may also help to explain the unusually hard 
X-ray spectrum \cite{g98}, if it has a low-mass ($\sim 10^{-20}-10^{-19}\,
\msun$) atmosphere \cite{u97}.

Strange stars have been proposed to model $\gamma$-ray bursters 
\cite{hpa91}, soft $\gamma$-ray repeaters \cite{cd98} and the bursting X-ray 
pulsar GRO J1744-28 \cite{c98}. But these models are generally speculative. 
In this work, we have suggested that \sax\ is a likely strange star 
candidate, by comparing its $M-R$ relation determined from X-ray observations
with the theoretical models of a neutron star and of a strange star.
If so, there will be very deep consequences for both the physics of strong
interactions and astrophysics.
But we point out that the available observational data are not sufficient 
and accurate enough to exclude the possibility that \sax\ could be a 
neutron star with anomalous magnetic fields. 
It has been suggested that strange stars could become unstable 
to $m=2$ bar mode \cite{cm92}.  Further observations of this signature in 
case of \sax\ will be of great interest.
 
We are grateful to Dr. G. B. Cook for providing the tabulated data for 
EOS UU of neutron stars, and the referees for critical comments. 
I. B. thanks Prof. B. Datta for valuable discussions  and helpful 
suggestions. X. L. was supported by National Natural Science Foundation of 
China and by the Netherlands Organization for Scientific Research (NWO).
J. D. and M. D. acknowledge partial support from Government of India
(SP/S2/K18/96) and FAPESP.

\vfil\eject

\end{document}